\documentclass{article}
\usepackage{spconf,amsmath,graphicx}

\usepackage{cite}
\usepackage{amsmath,amssymb,amsfonts}
\usepackage{algorithmic}
\usepackage{graphicx}
\usepackage{textcomp}
\usepackage{xcolor}
\def\BibTeX{{\rm B\kern-.05em{\sc i\kern-.025em b}\kern-.08em
    T\kern-.1667em\lower.7ex\hbox{E}\kern-.125emX}}
    
\usepackage[nohyperlinks, printonlyused,nolist]{acronym} %,
\usepackage{booktabs}
\usepackage{hhline}
\usepackage{multirow}

\usepackage{tikz}     
\usepackage{pgfplots}                %
\usepgfplotslibrary{groupplots}
\usetikzlibrary{plotmarks}
\usetikzlibrary{calc}
\usepgfplotslibrary{external}

\def\R{{\mathbb R}}

\usepackage{balance}
\usepackage{booktabs}
\usepackage{multirow}
\usepackage{hhline}
\usepackage{makecell}

\usepackage{amsmath,graphicx}
\usepackage{tikz}

\usepackage{pgfplots}
\usepgfplotslibrary{groupplots}

\usepackage{tikz}
\usetikzlibrary{shapes,arrows,calc,fit,automata,positioning}
\usepackage{amsmath,amsfonts,amssymb}
\usepackage{pdfpages}
\usetikzlibrary{intersections,backgrounds}
\usetikzlibrary{calc}
\usetikzlibrary{decorations.pathreplacing,calligraphy}

\let\OLDthebibliography\thebibliography
\renewcommand\thebibliography[1]{
  \OLDthebibliography{#1}
  \setlength{\parskip}{0pt}
  \setlength{\itemsep}{1pt}
}

\usepackage{paralist}
\usepackage[hidelinks]{hyperref}
\title{\MakeUppercase{Multi-Microphone Speaker Separation by Spatial Regions}}
\name{Julian Wechsler\textsuperscript{$\dagger$\thanks{$^\dagger$Corresponding author: julian.wechsler@audiolabs-erlangen.de.}}, Srikanth Raj Chetupalli, Wolfgang Mack, Emanu{\"e}l A.~P. Habets}

\address{
International Audio Laboratories Erlangen$^*$\thanks{$^*$A joint institution of the Friedrich-Alexander-Universit\"{a}t Erlangen-N\"{u}rnberg (FAU) and Fraunhofer IIS.}, Am Wolfsmantel 33, 91058 Erlangen, Germany
}
\begin{document}
\ninept

\maketitle
\begin{abstract}
We consider the task of region-based source separation of reverberant multi-microphone recordings.
We assume pre-defined spatial regions with a single active source per region.
The objective is to estimate the signals from the individual spatial regions as captured by a reference microphone while retaining a correspondence between signals and spatial regions.
We propose a data-driven approach using a modified version of a state-of-the-art network, where different layers model spatial and spectro-temporal information.
The network is trained to enforce a fixed mapping of regions to network outputs.
Using speech from LibriMix, we construct a data set specifically designed to contain the region information.
Additionally, we train the network with permutation invariant training.
We show that both training methods result in a fixed mapping of regions to network outputs, achieve comparable performance, and that the networks exploit spatial information.
The proposed network outperforms a baseline network by $1.5$~dB in scale-invariant signal-to-distortion ratio.
\end{abstract}
\section{Introduction}
\label{sec:intro}

\acresetall
When capturing a mixture of multiple speakers with an array of microphones, the signals contain information related to the speakers' identities and their positions.
The speakers' identities, thereby, are encoded in the spectral structure of the individual microphone signals. The positions are implicitly represented in time differences of arrival and attenuation differences of the received speech signals caused by the relative positioning of the source and array. This information can be exploited to separate the speakers from each other, as required, e.g., by automatic speech recognition systems \cite{espnet}, hearing aids \cite{HA}, or communication systems \cite{vincent2018audio}. 

Speaker separation systems are typically implemented as \acp{DNN} that estimate a mask for each speaker from a feature representation of the microphone signal(s). Subsequently, the mask is applied element-wise to a feature representation of the microphone signal(s) to estimate the features of the individual speakers.
Each speaker's features are transformed back to the time domain such that they can be processed separately.
During training the \ac{DNN}, the so-called permutation problem occurs.
As all speakers are instances of the class ``speaker'', there are multiple possibilities (speaker permutations) to assign the masks to the individual speakers.
A prominent solution to address the permutation problem is deep clustering, where mixture features are mapped to vectors \cite{DC,DAN,MCDC,USDC}.
The permutation problem is addressed by using a vector-affinity-based loss.
An alternative solution is \ac{PIT} \cite{PIT,UPIT,GammatonePIT,TDUPIT,TFGRID,AmbiSep}, where a reconstruction loss of all permutations is computed, and only the smallest loss amongst the permutations is used to update the \ac{DNN}.
However, \ac{PIT} does not scale well with the number of speakers. 
In \cite{Taherian2022a,Taherian2022b,subramanian2022,xu2022learning, coneofsilence}, the authors avoided the permutation problem by consistently selecting a permutation based on the relative speaker positions (distance or angle) w.r.t. the microphone array. Note that fixing the permutation based on relative positions does not enable to infer the speaker positions from the estimated outputs. 
To extract a single source from a defined target area,  \cite{markovic22_interspeech, Tesch2023} train \acp{DNN} that exploit location information implicitly given in the \ac{DNN} inputs.  Alternative approaches \cite{Thiergart_ISF, Sivasankaran} provide the target \ac{DOA} explicitly to the separation system such that it extracts the respective source. 

In some application scenarios, the control rights of speakers can be defined by their positions.
Consider, for example, a car environment where only the driver should be able to control specific applications with their voice (e.g., an autopilot system).
Single-source extraction methods \cite{markovic22_interspeech, Tesch2023} may not be suitable for such scenarios, as all sources are required, whereas methods that are explicitly provided with the target \acp{DOA} as input may exhibit problems with spatially extended or angularly overlapping regions \cite{Thiergart_ISF, Sivasankaran}. 

To relate a \ac{DNN} output to a specific spatial region, the main contributions of this paper are 1) to introduce the task of multi-source separation with sources located in fixed pre-defined regions, where the desired output signals are assigned to associated regions by enforcing a one-to-one correspondence between regions and network outputs, 2) to propose a training data generation scheme with a task-based spatial data distribution, i.e., to construct the training data such that the speaker positions are restricted to pre-defined spatial regions, and 3) to apply the recently proposed triple-path structure \cite{AmbiSep,TPRNN} for multi-source separation in the time domain, where individual \ac{DNN} layers model inter-channel, short-term, and long-term information.

\section{Problem Formulation}
\label{sec:ProbForm}

Consider an in-car scenario where the speech of $S \in \mathbb{N}^+$ speakers is captured by a \ac{ULA} with $M \in \mathbb{N}^+$ microphones mounted, e.g., on the rear mirror.
The set of positive integers is given by $\mathbb{N}^+$.
The $m^\mathrm{th}\in \{1,\ldots,M\}$ discrete time-domain microphone signal $\mathbf{y}_m \in \mathbb{R}^{T\times 1}$ is a mixture of the $S$ reverberant speech signals $\mathbf{x}_m^{(s)} \in \mathbb{R}^{T\times 1}$, i.e., 
\begin{equation}
    \mathbf{y}_m = \sum_{s=1}^{S} \mathbf{x}_m^{(s)}, 
\end{equation}
where $s \in \{1, \ldots, S\}$ is the speaker index, $T \in \mathbb{N}^+$ is the length of the signals in time samples, and $\mathbb{R}$ is the set of real numbers.
In a car, the speaker positions can be restricted to $R \in \mathbb{N}^+$ distinct regions with index $r \in \{1, \ldots, R\}$ (e.g., the passenger seats). %, composing three distinct spatial regions.  
We denote the set of speakers from region $r$ by $\mathcal{S}^{(r)}$ and define the mixture in region $r$ by
\begin{equation}
\mathbf{x}^{(r)}_{m} = \sum_{s \in \mathcal{S}^{(r)}}
\mathbf{x}^{(s)}_{m}.
\end{equation}

In the present work, we tackle the problem of separating speech sources belonging to different pre-defined, fixed spatial regions. In addition, each output signal is associated with one region. This association allows to directly identify, e.g., the driver, such that region-specific control rights can be implemented in a human-machine interface. This problem can be understood as a specific \ac{MIMO} signal estimation problem. 
More formally, the objective is to obtain estimates $\mathbf{\widehat{x}}^{(r)}_{m_\text{ref}}$ of the reverberant speech mixtures per region at a reference microphone $m_\text{ref}$ based on the $M$ microphone signals $\mathbf{y}_m$. In this work, we assume a single active speaker per region (i.e., $R=S$).

\section{Separation Architecture}
\label{sec:ambisep}

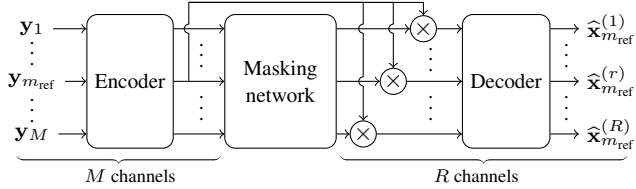
\begin{figure}[t]
    \centering
    \tikzstyle{line} = [draw, ->]
    \usetikzlibrary{decorations.pathreplacing,calligraphy}
    \begin{tikzpicture}
        \node[draw, align=center, rounded corners, minimum height=0.7in] (Enc) at (0,0.2) {\footnotesize Encoder};
        \node[draw, text width = 0.5in, align=center, rounded corners, minimum height=0.7in] (Masker) at (2.0,0.2) {\footnotesize Masking network};
        \node[draw, align=center, rounded corners, minimum height=0.7in] (Dec) at (5.0,0.2) {\footnotesize Decoder};

        \node[] (y0) at (-1.3,0.9cm) {\footnotesize ${\bf y}_1$};
        \node[]  at (-1.3,0.68) {\footnotesize $\vdots$};
        \node[] (y1) at (-1.3,0.2) {\footnotesize ${\bf y}_{m_\text{ref}}$};
        \node[]  at (-1.3,-0.08) {\footnotesize $\vdots$};
        \node[] (yM) at (-1.3,-0.5cm) {\footnotesize ${\bf y}_M$};
        
        \draw[->] (y0) -- ( y0 -| Enc.west);
        \draw[->] (y1) -- ( y1 -| Enc.west);
        \draw[->] (yM) -- ( yM -| Enc.west);
        
        \path [line,name path=EncToMask0] ([yshift=0.7cm] Enc.east) -- ([yshift=0.7cm] Masker.west);
        
        \draw[->] ([yshift=0cm] Enc.east) -- ([yshift=0cm] Masker.west);
        
        \draw[->] ([yshift=-0.7cm] Enc.east) -- ([yshift=-0.7cm]  Masker.west);
        \node[]  at (0.95,-0.05) {\footnotesize $\vdots$};
        \node[]  at (0.95,0.65) {\footnotesize $\vdots$};

        \node[draw,circle,inner sep=0.005cm] (m0) at (3.9,0.9cm) {$\times$};
        \node[draw,circle,inner sep=0.005cm] (m1) at (3.5,0.2cm) {$\times$};
        \node[draw,circle,inner sep=0.005cm] (m2) at (3.1,-0.5cm) {$\times$};

        \path[line, name path=MaskToM0] ([yshift=0.7cm] Masker.east) -- (m0.west);
        \path[line, name path=MaskToM1] ([yshift=0cm] Masker.east) -- (m1.west);
        \draw[->] ([yshift=-0.7cm] Masker.east) -- (m2.west);
        
        % Draw path with "skip" of line crossing from Enc1 upwards
        \coordinate (S)  at ([xshift=0.2cm, yshift=0cm] Enc.east);
        \coordinate (UP) at ([xshift=0.2cm, yshift=1.05cm] Enc.east);
        
        \path [name path=StoUP] (S) -- (UP);
        \path [name intersections={of=EncToMask0 and StoUP}];
        \coordinate (C)  at (intersection-1);
        \path [name path=circle] (C) circle(0.6mm);
        
        \path[name intersections={of=circle and StoUP}];
        \coordinate (I1)  at (intersection-1);
        \coordinate (I2)  at (intersection-2);
        
        \draw (S) -- (I2);
        \draw[-] (I1)  -- (UP);

        \draw (I2) arc (270:90:0.6mm);
        
        \draw[->] (UP) -| (m0.north);
        
        % Draw path with "skip" of line crossing from "upwards" position to mask multiplication of channel 1
        \path[name path=UPtoM1] (UP) -| (m1.north);
        \path[name intersections={of=UPtoM1 and MaskToM0}];
        \coordinate (Cprime)  at (intersection-1);
        \path [name path=circle2] (Cprime) circle(0.6mm);
        \path[name intersections={of=circle2 and UPtoM1}];
        \coordinate (I12)  at (intersection-1);
        \coordinate (I22)  at (intersection-2);
        
        \draw (UP) -| (I12);

        \draw (I22) arc (270:90:0.6mm);
        
        \draw[->] (I22) -| (m1.north);
        
        % Draw path with "skip" of line crossing from "upwards" position to mask multiplication of channel 2
        \path[name path=UPtoM2] (UP) -| (m2.north);
        \path[name intersections={of=UPtoM2 and MaskToM0}];
        \coordinate (Cdoubleprime)  at (intersection-1);
        \path [name path=circle3] (Cdoubleprime) circle(0.6mm);
        \path[name intersections={of=circle3 and UPtoM2}];
        \coordinate (I13)  at (intersection-1);
        \coordinate (I23)  at (intersection-2);
        
        \draw (UP) -| (I13);
        
        \draw (I23) arc (270:90:0.6mm);
        
        \path[name path=I23toM2] (I23) -> (m2.north);
        \path[name intersections={of=I23toM2 and MaskToM1}];
        \coordinate (Ctripleprime)  at (intersection-1);
        \path [name path=circle4] (Ctripleprime) circle(0.6mm);
        \path[name intersections={of=circle4 and I23toM2}];
        \coordinate (I14)  at (intersection-1);
        \coordinate (I24)  at (intersection-2);
        
        \draw (I23) -- (I14);
        \draw (I24) arc (270:90:0.6mm);
        \draw[->] (I24) -> (m2.north);
        
        % \draw[->, blue] (S) -- ([xshift=0.2cm, yshift=0.9cm] Enc.east) -| (m1.north);
        
        % \draw[->, blue] (S) -- ([xshift=0.2cm, yshift=0.9cm] Enc.east) -| (m2.north);
        
        \draw[->] (m0.east) -- ( m0.east -| Dec.west);
        \draw[->] (m1.east) -- (m1.east -| Dec.west);       
        \draw[->] (m2.east) -- ( m2.east -| Dec.west);

        \node[] (s0) at (6.4,0.9cm) {\footnotesize ${\bf \widehat x}_{m_\text{ref}}^{(1)}$};
        \node[] (s1) at (6.4,0.2cm) {\footnotesize ${\bf \widehat x}_{m_\text{ref}}^{(r)}$};
        \node[] (s2) at (6.4,-0.5cm) {\footnotesize${\bf \widehat x}_{m_\text{ref}}^{(R)}$};
        \node[] at (5.8,-0.05cm) {\footnotesize $\vdots$};
        \node[] at (4.0,-0.05cm) {\footnotesize $\vdots$};
        
        \node[] at (5.8,0.65cm) {\footnotesize $\vdots$};
        \node[] at (4.0,0.6cm) {\footnotesize $\vdots$};
        
        \draw[->] ([yshift=0.7cm] Dec.east) -- (s0);
        \draw[->] ([yshift=0cm]Dec.east) -- (s1);
        \draw[->] ([yshift=-0.7cm] Dec.east) -- (s2);
        
        \draw [decorate,
    decoration = {calligraphic brace,mirror}] (-1.5,-0.8) --  (1.2,-0.8);
    
        \draw [decorate,
    decoration = {calligraphic brace,mirror}] (2.8,-0.8) --  (6.6,-0.8);
    
        \node at (0.,-1.05) {\scriptsize $M$ channels};
        
        \node at (4.6,-1.05) {\scriptsize $R$ channels};

    \end{tikzpicture}
    \vspace{-0.5cm}
    \caption{\ac{MIMO} time-domain source-separation framework.}
    \label{fig:MaskingInTFD}
    \vspace{-0.25cm}
\end{figure}

The employed \ac{DNN} is a modified version of the recently proposed AmbiSep \cite{AmbiSep}, referred to as \ac{SpaRSep}, where masking in a learned \ac{TF} domain \cite{TaSNet} is used for separation  (see Figure~\ref{fig:MaskingInTFD}).
The encoder performs short-time analysis and converts the $M$ time-domain inputs to sub-sampled, positive \ac{TF} representations, i.e.,
\begin{equation}
    \mathbf{Y}_m = \text{Encoder}\left(\mathbf{y}_m\right),~~~\mathbf{Y}_m\in\R^{N\times F}_+,
\end{equation}
where $N$ is the number of time frames, and $F$ is the number of features.
The masking network generates one positive real-valued mask ${\bf M}_r$ for each source region $r$, which is multiplied with the encoder output of the reference microphone channel, $\mathbf{Y}_\text{ref}$.
The masked representations are converted back to the time domain by the decoder. 

\textbf{Encoder-Decoder:}
A $1$D convolution layer of $F$ filters with a kernel-size equivalent to $1$~ms duration and $50\%$ stride, followed by \ac{ReLU} activation, is used as the encoder.
The decoder comprises a transpose-$1$D convolution layer with the same parameters as the encoder.
The encoder and decoder are applied independently to the input and output channels, respectively.
Note that the weights are shared across the channels.

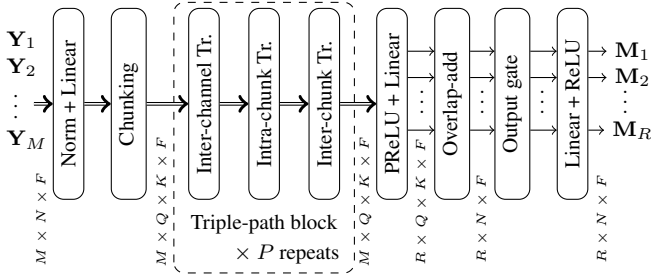
\begin{figure}[t]
    \centering
    \begin{tikzpicture}
        \node[text width=0.1in, align=center] (In) at (-0.3,0) {\footnotesize ${\bf Y}_1$ ${\bf Y}_2$ $\vdots$ ${\bf Y}_M$ $\phantom{{\bf Y}_M}$};
        \node[draw, align=center, rounded corners, rotate=90, minimum width=1.in] (LLN) at (0.4,0) {\footnotesize Norm + Linear};
        \node[draw, align=center, rounded corners, rotate=90, minimum width=1.in] (CH) at (1.2,0) {\footnotesize Chunking};
        \node[draw, align=center, rounded corners, rotate=90, minimum width=1.in] (Itra) at (2.2,0) {\footnotesize Inter-channel Tr.};
        \node[draw, align=center, rounded corners, rotate=90, minimum width=1.in] (Iter) at (3.0,0) {\footnotesize Intra-chunk Tr.};
        \node[draw, align=center, rounded corners, rotate=90, minimum width=1.in] (IterC) at (3.8,0) {\footnotesize Inter-chunk Tr.};
        \draw[dashed, rounded corners] (1.8,-2.25) rectangle (4.2,1.35);
        \node[align=center] at (3.0,-1.6) {\footnotesize Triple-path block};
        \node[align=center] at (3.3,-2) {\footnotesize $\times~P$ repeats};
        \node[draw, align=center, rounded corners, rotate=90, minimum width=1.in] (C2) at (4.7,0) {\footnotesize PReLU + Linear};
        \node[draw, align=center, rounded corners, rotate=90, minimum width=1.in] (OVA) at (5.5,0) {\footnotesize Overlap-add};
        \node[draw, align=center, rounded corners, rotate=90, minimum width=1.in] (GO) at (6.3,0) {\footnotesize Output gate};
        \node[draw, align=center, rounded corners, rotate=90, minimum width=1.in] (LRU) at (7.1,0) {\footnotesize Linear + ReLU};
        \node[] (m0) at (7.9,0.7) {\footnotesize ${\bf M}_1$};
        \node[] (m1) at (7.9,0.35) {\footnotesize ${\bf M}_2$};
        \node[] (m2) at (7.9,-0.35) {\footnotesize ${\bf M}_R$};
        \node[rotate=90] (ms) at (7.8,0.05) {\footnotesize $\dots$};

        \draw[->,double] (In.east) -- (LLN);
        \draw[->,double] (LLN) -- (CH);
        \draw[->,double] (CH) -- (Itra);
        \draw[->,double] (Itra) -- (Iter);
        \draw[->,double] (Iter) -- (IterC);
        \draw[->,double] (IterC) -- (C2);

        \draw[->] ([yshift=0.7cm]C2.south) -- ([yshift=0.7cm]OVA.north);
        \draw[->] ([yshift=0.35cm]C2.south) -- ([yshift=0.35cm]OVA.north);
        \draw[->] ([yshift=-0.35cm]C2.south) -- ([yshift=-0.35cm]OVA.north);
        \node[rotate=90] (ms) at (5.1,0.05) {\footnotesize $\dots$};
        
        \draw[->] ([yshift=0.7cm]OVA.south) -- ([yshift=0.7cm]GO.north);
        \draw[->] ([yshift=0.35cm]OVA.south) -- ([yshift=0.35cm]GO.north);
        \draw[->] ([yshift=-0.35cm]OVA.south) -- ([yshift=-0.35cm]GO.north);
        \node[rotate=90] (ms) at (5.9,0.05) {\footnotesize $\dots$};
        
        \draw[->] ([yshift=0.7cm]GO.south) -- ([yshift=0.7cm]LRU.north);
        \draw[->] ([yshift=0.35cm]GO.south) -- ([yshift=0.35cm]LRU.north);
        \draw[->] ([yshift=-0.35cm]GO.south) -- ([yshift=-0.35cm]LRU.north);
        \node[rotate=90] (ms) at (6.7,0.05) {\footnotesize $\dots$};

        \draw[->] (LRU.south |- m0.west) -- (m0.west);
        \draw[->] (LRU.south |- m1.west) -- (m1.west);
        \draw[->] (LRU.south |- m2.west) -- (m2.west);

        \node[rotate=90] at (0.05,-1.53) {\tiny $M\times N\times F$};
        \node[rotate=90] at (1.65,-1.3) {\tiny $M\times Q\times K\times F$};
        \node[rotate=90] at (4.35,-1.3) {\tiny $M\times Q\times K\times F$};
        \node[rotate=90] at (5.1,-1.3) {\tiny $R\times Q\times K\times F$};
        \node[rotate=90] at (5.9,-1.53) {\tiny $R\times N\times F$};
        \node[rotate=90] at (7.5,-1.53) {\tiny $R\times N\times F$};   
        % \node[align=center] at (4,-2.6) {\footnotesize  $N$: \#Frames, $Q$: \#Chunks, $K$: Chunk size,   };
    \end{tikzpicture}
    \vspace{-0.5cm}
    \caption{Masking network architecture of \ac{SpaRSep} (after \cite{AmbiSep}).}
    \label{fig:MaskingNetwork}  
    \vspace{-0.45cm}
\end{figure}

\textbf{Masking network:}
Figure \ref{fig:MaskingNetwork} shows a block diagram of the masking network.
The input multi-channel \ac{TF} representation is first passed through a LayerNorm layer \cite{layernorm} followed by a linear layer, and is then segmented into $Q$ chunks of size $K$ with $50\%$ overlap.
The chunked \ac{TF} representation is fed to  $P$ concatenated \acp{TPB}.
Each \ac{TPB} comprises three transformer encoder layers \cite{Vaswani2017Attention}, one inter-channel, one intra-chunk, and one inter-chunk transformer.
The inter-channel transformer models the relationships across the microphone channels (i.e., spatial information) in the input at each time frame.
The intra-chunk transformer models the short-time relations in the input, while the inter-chunk transformer models the global relations across the chunks.

The output of the last \ac{TPB} is passed through a \ac{PReLU} layer.
As opposed to \cite{AmbiSep}, the signals are flattened along the channel dimension before being fed to a linear layer to compute $R$ outputs.
The chunking operation is inverted using overlap-add, and the corresponding output is passed through an output gate.
The gated output is then passed through a linear layer with ReLU activation to predict positive-valued masks.

All layers in the architecture, except for the inter-channel transformers, process the inputs independently across the channels, making the architecture amenable to a parallel implementation across the channels. 
An important feature of the architecture is that the number of parameters in the model is independent of the number of input microphones $M$ (which is the sequence dimension in the inter-channel transformer) and sources $S$.
Only the output size of the linear layer after all \acp{TPB} depends on the number of regions $R$.

\section{Proposed Region-Based Separation}
\label{sec:ProposedMethod}

\begin{figure}
    \centering
    \scalebox{0.8}{
	\begin{tikzpicture}% auto for automatic positioning
		 \node (regions) {\includegraphics[scale=0.1]{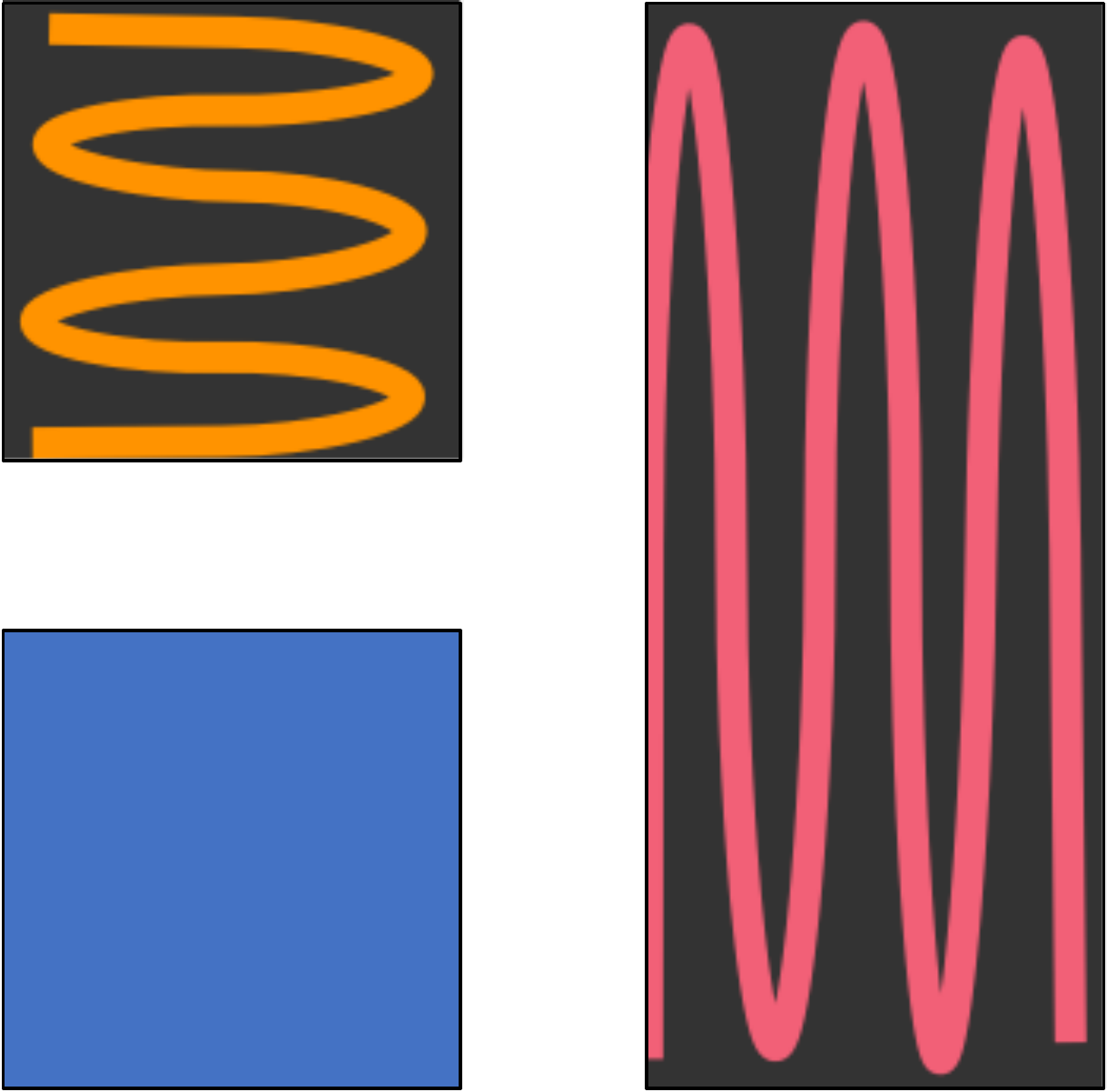}};
    \node[below = 0.9cm of regions.center](regioncaption){3 Regions};

		 \node[right= of regions, yshift=0.3cm, xshift=-0.8cm] (PIT1) {\includegraphics[scale=0.1]{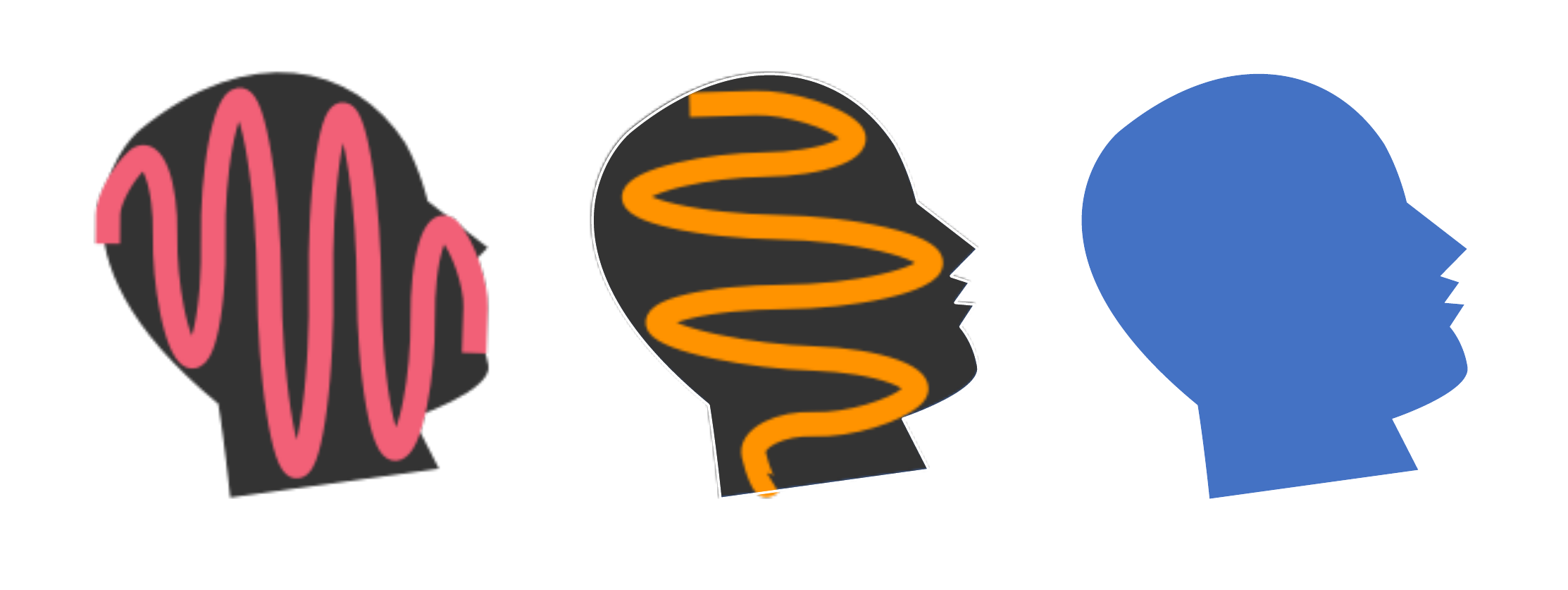}};
    \node[right = 1.2cm of PIT1.center](PIT1cap){Perm. 1};
    		 \node[below= 0.2 cm of PIT1] (PIT2) {\includegraphics[scale=0.1]{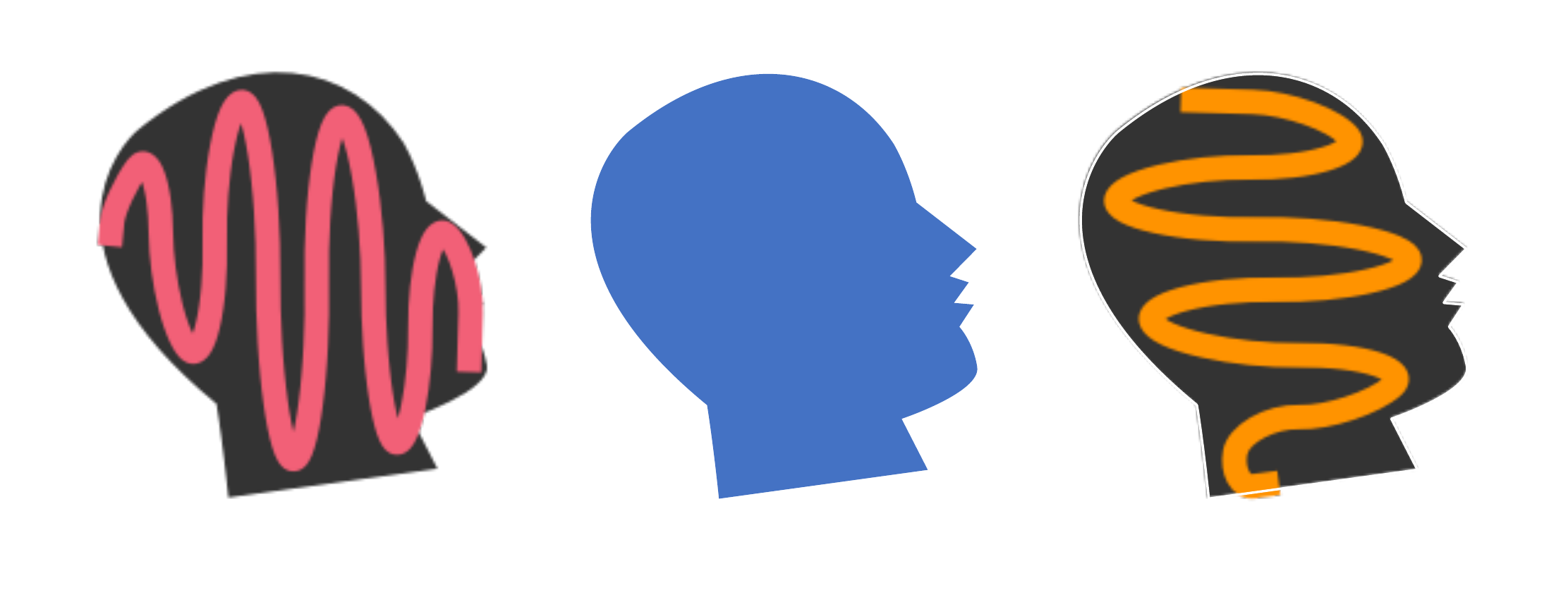}};
    \node[right= 1.2cm of PIT2.center](PIT2cap){Perm. 2};

\node[fill=gray,circle,draw,scale=0.7, left = 1cm of regions] (mic1) {};
        \draw ([yshift=0.15cm]mic1.east) -- ([yshift=-0.15cm]mic1.east);
       	
       	\node[fill=gray,circle,draw,scale=0.7, below=0.05 cm of mic1] (mic2) {};
        \draw ([yshift=0.15cm]mic2.east) -- ([yshift=-0.15cm]mic2.east);
               	\node[fill=gray,circle,draw,scale=0.7,above=0.05cm of mic1] (mic3) {};
        \draw ([yshift=0.15cm]mic3.east) -- ([yshift=-0.15cm]mic3.east);
            \node[below= 0.9cm of mic1.center] (ArrayCaption) {Array};
    
\draw[rectangle,label=Room]($(mic1.center)+(-1cm,-1.5cm)$) rectangle ++(3.5cm, 2.2cm) ;
	\end{tikzpicture}}
 \vspace{-0.3cm}
    \caption{Illustration of a room with a microphone array and three regions, each with one active speaker (left). On the right, two of the possible six permutations (e.g., when using \ac{PIT}) are illustrated. Speaker-region correspondence is illustrated by the colors. }
    \label{fig:PIT}
    \vspace{-0.45cm}
\end{figure}
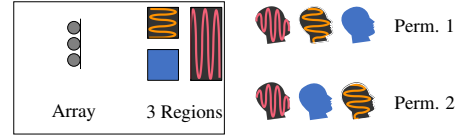

To preserve the region information on the $R$ separated signals $\widehat{\mathbf{x}}^{(r)}_{m_\text{ref}}$, we propose to assign each of the $R$ output channels to a specific (different) region $r$.
We propose to implement that assignment in two steps.
First, we propose to generate training data where the speaker positions are limited to the respective regions (one speaker per region).
We train \ac{SpaRSep} to maximize the commonly used \ac{SI-SDR} \cite{SISDR} (also, cf. \cite{AmbiSep}).
Typically, speaker separation systems are trained with \ac{PIT}, where \ac{DNN} updates are based on the speaker permutation that yields the lowest loss.
An illustration of permutations for three speakers is given in Figure~\ref{fig:PIT}.
As the proposed training data generation yields data where the speaker permutation can be fixed according to spatial regions, \ac{PIT} could converge to a permutation that preserves the region information.
In other words, the estimated speech signals may consistently be ordered according to the regions.
However, it is unclear how \ac{PIT} determines the order and whether \ac{PIT} will converge to a region-dependent permutation.
Consequently, we propose to consistently assign each region to a specific \ac{DNN} output during training.
That way, we ensure that the \ac{DNN} is trained with a region-dependent permutation.
Additionally, the computational load during training is reduced compared to \ac{PIT}, most notably when the number of regions increases:
While PIT scales factorially, the fixed mapping scales linearly with the number of regions $R$.
For example, when training with a fixed mapping for three regions, we always update the \ac{DNN} with one of the six possible permutations, e.g., one shown in Figure~\ref{fig:PIT}.

\section{Experimental Setup}
\label{sec:ExSet}

This section briefly describes the simulation of the data sets and gives the parameters of the employed \ac{DNN} architecture.

\textbf{Datasets:}
To model an in-car environment, we simulated \acp{RIR} for a car-sized shoe-box room (3~m,   2~m, 1.5~m) with typical in-car reverberation times from $0.05{-}0.1$ seconds in steps of 0.01 \cite{carT60} (slightly exceeding values from \cite{carT60} to accommodate for different materials like glass ceilings).
The room contains a \ac{ULA} with $M=3$ microphones and $8$~cm inter-microphone spacing where the central microphone was placed at (0.5~m, 1~m, 1~m).
Additionally, there are three distinct cuboid regions where speakers can be active (referred to as `driver' (D), `co-driver' (C), `backseats' (B)). The 'driver' and `co-driver' region-centers (cubes with an edge length of 0.5~m) were placed at (1.25~m, 0.5~m, 1~m) and (1.25~m, 1.5~m, 1~m), respectively. The `backseats' region consists of a cuboid region three times the size of one individual front region, placed 1~m behind the `driver'/`co-driver' regions (see Figure~\ref{fig:scatter-plot}). Fifty points were randomly sampled from the two smaller regions (D and C), 150 from the bigger one (B). Per region, the source positions were split into three disjoint subsets (60\%, 20\%, 20\%) as basis for training, validation, and test sets, respectively.

Based on source and microphone positions, \acp{RIR} were simulated using \cite{RIRGenerator}.
They were convolved with speech according to the Libri3Mix \cite{LibriMix} description in the configurations `mix\_clean' and `min' to obtain the reverberant speech signals at the microphones.
The sampling frequency was 16~kHz, and the length of training and validation files was four seconds. To form a single training/validation/test example, three reverberant speech signals, one per region, were added. We refer to that test set as \ac{FC}. For the test, we additionally created a set with simulated microphone self-noise by adding spatio-temporal white Gaussian noise with a signal-to-noise ratio $\in [20,30]$~dB w.r.t. all reverberant speakers.  We refer to this test set as \ac{WN}. To show that the \ac{DNN} can assign the speakers to the regions even if not all regions have an active speaker simultaneously, we create a third test set referred to as \ac{PO}. There, the speakers start speaking successively with a delay of two seconds. The speaker order is determined randomly. Overall, the training, validation, and each test set consist of  9'300, 3'000, and 3'000 samples, respectively.

\textbf{Model Details:}
We trained two copies of the \ac{SpaRSep} architecture as described in Section~\ref{sec:ambisep} for $100$ epochs using an initial learning rate of $15\mathrm{e}{-5}$ and a batch size of one sample.
One copy was trained using \ac{PIT} (as in \cite{AmbiSep}), and one copy was trained using the fixed mapping as motivated and proposed in Section~\ref{sec:ProposedMethod}, both times employing \ac{SI-SDR} as loss function.
For the variant without \ac{PIT}, we enforced the source from the `driver region' be mapped to output $1$, `co-driver' to $2$, `backseats' to $3$.
The encoder was set to encode the waveform signals using a window length of $16$ samples and a hop size of $8$ samples, outputting a representation with $F=128$ features.
The resulting features were chunked with a chunk size of $K=250$ and passed through a concatenation of $P=4$ \acp{TPB}.
The model comprises approximately $4.2$ million parameters.

\section{Performance Evaluation}
\label{sec:PerformanceEval}

\begin{figure}
    \centering
    \scalebox{0.8}{
	\begin{tikzpicture}% auto for automatic positioning
		    \node (all) {\includegraphics[width=9cm,height=6cm]{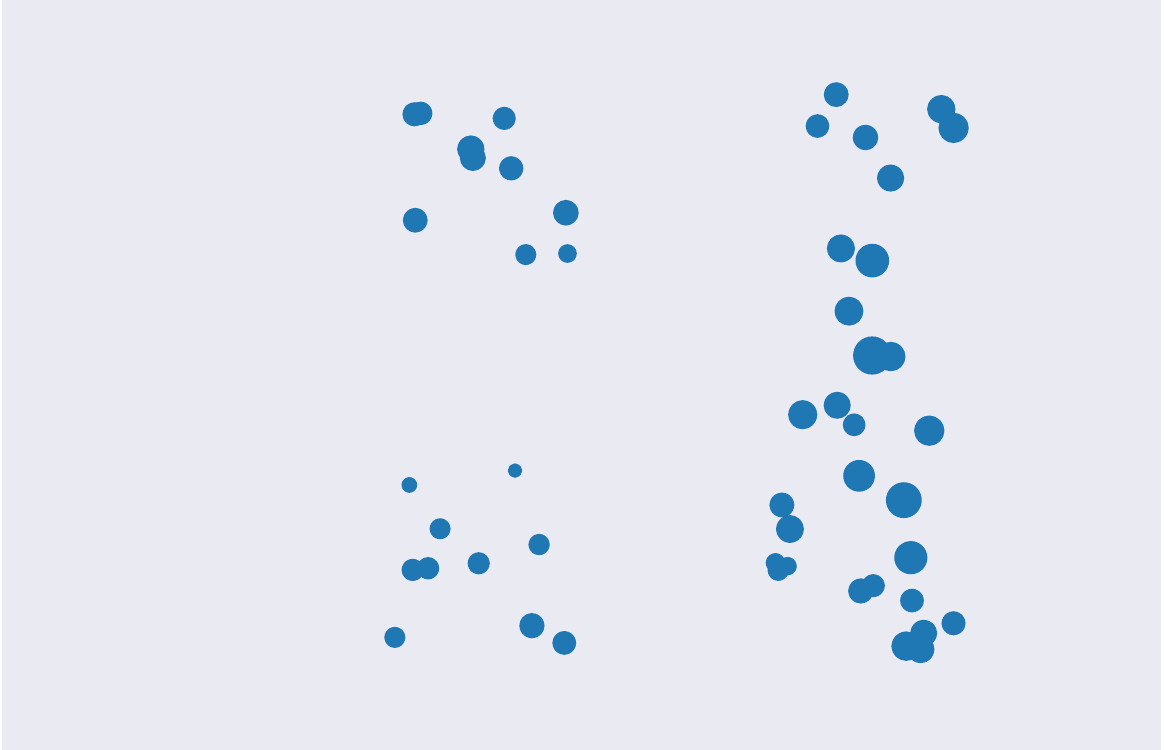}};
		%%% Places 
 		\node[draw,rectangle,minimum width=9cm,minimum height=6cm] (room) at (0,0) {};
 		\node[draw,rectangle,minimum width=1.5cm, minimum height=1.5cm] (Codriver) at (-0.75cm, 1.5cm) {};
   
        \node[below=0.2cm of Codriver] (Codriver Caption) {Co-Driver Seat};

		\node[above= 0.1cm of Codriver] (sizeseat){0.5~m};	
		\node[right= 0.1cm of Codriver] (sizeseat){\rotatebox{-90}{0.5~m}};	

 		% Calligraphic brace
		\draw [decorate,decoration = {brace}] (Codriver.north west) --  (Codriver.north east);
		\draw [decorate,decoration = {brace}] (Codriver.north east) --  (Codriver.south east);
		
		\node[draw,rectangle,minimum width = 1.5cm, minimum height=1.5cm](Driver) at (-0.75cm,-1.5cm) {};
		    \node[below=0.2cm of Driver] (Driver Caption) {Driver Seat};
   
	\node[draw,rectangle,minimum width = 1.5cm, minimum height=4.5cm] (Back) at (2.25cm,-0cm) {};
		\node[above= 0.1cm of Back] (sizeseat){0.5~m};	
		\node[right= 0.1cm of Back] (sizeseat){\rotatebox{-90}{1.5~m}};	 		
 		\draw [decorate,decoration = {brace}] (Back.north west) --  (Back.north east);
		\draw [decorate,decoration = {brace}] (Back.north east) --  (Back.south east);

      \node[below=0.2cm of Back] (Back Caption) {Back Seats};
   
       % \draw[->] (-3.8cm,-2.9cm) -- (-3.8cm, -2.4cm);
        %\draw[->] (-3.8cm,-2.9cm) -- (-3.3cm, -2.9cm);
        %\node (xlabel) at (-3.2cm, -2.8cm) {width};

 		\node[below=0.05cm of room.south](width){Length: 3~m};
 			\node[left=0.05cm of room.west](width){\rotatebox{90}{Width: 2~m}};
 	
 		%% Microphones
		\node[fill=gray,circle,draw,scale=0.7] (mic1) at (-3cm,0cm){};
        \draw ([yshift=0.15cm]mic1.east) -- ([yshift=-0.15cm]mic1.east);
       	
       	\node[fill=gray,circle,draw,scale=0.7] (mic2) at (-3.0cm,-0.24cm){};
        \draw ([yshift=0.15cm]mic2.east) -- ([yshift=-0.15cm]mic2.east);
               	\node[fill=gray,circle,draw,scale=0.7] (mic3) at (-3.0cm,0.24cm){};
        \draw ([yshift=0.15cm]mic3.east) -- ([yshift=-0.15cm]mic3.east);
            \node[below=0.75cm of mic3] (ArrayCaption) {Array};
            
            \coordinate (up) at ([xshift=0cm, yshift=-0.75cm] Back.north east);
            \coordinate (down) at ([xshift=0cm, yshift=0.75cm] Back.south east);
            
           \draw[-,dotted] (mic1.center) -- (up);  \draw[-,dotted] (mic1.center) -- (down);
           
           \draw[-,dotted] (mic1.center) -- (Back.north west);
           \draw[-,dotted] (mic1.center) -- (Back.south west);

	\end{tikzpicture}}
    \vspace{-0.25cm}
    \caption{Illustration of the simulated room with points at the speakers' positions. The point sizes are proportional to the mean \acs{SI-SDR} improvement per position. The room proportions (size, zones, and positions) are true to scale.}
    \label{fig:scatter-plot}
    \vspace{-0.5cm}
\end{figure}
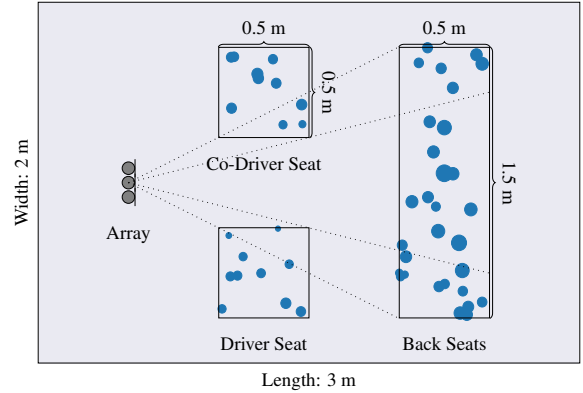

\begin{table*}[tbp]
    \caption{Mean \acs{SI-SDR} and \acs{SI-SDRi} results comparing training using \ac{PIT} and the proposed fixed-mapping training, evaluated on the test sets as described in Section~\ref{sec:ExSet}. All values are specified in dB. We report the metric as `\acs{SI-SDR} (\acs{SI-SDRi})'.}
    \vspace{-0.3cm}
	\begin{center}
	\resizebox{\textwidth}{!}{
	    \begin{tabular}{c c ccccc ccccc}
		\toprule
		\multicolumn{2}{c}{\multirow{2}{*}{}}&\multicolumn{4}{c}{Training with \acs{PIT}}&\multicolumn{4}{c}{Training with Proposed Fixed Mapping}\\
		& &average&`driver' &`co-driver'&`backseats'&average&`driver' &`co-driver'&`backseats'\\
		\hhline{~~----||----}
		{\makecell{MC ConvTasNet \cite{MC-ConvTasNet}}}& \acs{FC}&13.0\,(16.8) &15.0\,(16.3) &14.7\,(16.9) &\phantom{1}9.4\,(17.2) &- &- &- &-\\[0.075cm]
        \hhline{----------}\\[-0.25cm]
		\multirow{3}{*}{\ac{SpaRSep}}& \acs{FC}&14.6\,(18.3) &16.4\,(17.7) &16.4\,(18.4) &11.1\,(18.9) &14.5\,(18.3) &16.3\,(17.6) &16.1\,(18.3) &11.1\,(18.9)\\
		& \acs{WN}&13.4\,(17.2) &15.4\,(16.8) &15.3\,(17.5) &\phantom{1}9.6\,(17.4) &13.5\,(17.2) &15.5\,(16.8) &15.4\,(17.5) &\phantom{1}9.6\,(17.4)\\
		& \acs{PO}&19.1\,(22.9) &21.0\,(22.3) &20.8\,(23.0) &15.6\,(23.4) &19.7\,(23.5) &21.5\,(22.7) &21.2\,(23.4) &16.5\,(24.3)\\
		\bottomrule
    	\end{tabular}
    	}
    	\label{tab:performance_comparison}
	\end{center}
    \vspace{-1.0cm}
\end{table*}

\begin{table}[tbp]
    \caption{Analysis of the output permutation of AmbiSep for the 3'000 samples for every test set as described in Section~\ref{sec:ExSet}, comparing training using \ac{PIT} and the proposed fixed-mapping training. Both training runs converge to a fixed mapping (`majority'). We report the number of correct classifications and confusions between `driver' (D), `co-driver' (C), and `backseats' (B). At least one speaker was always assigned to the correct output.}
    \vspace{-0.2cm}
	\begin{center}
	    \begin{tabular}{r ccc ccc}
		\toprule
		\multirow{2}{*}{}&\multicolumn{3}{c}{\acs{PIT}}&\multicolumn{3}{c}{Proposed Fixed Mapping}\\
		& \acs{FC}& \acs{WN}& \acs{PO}&\acs{FC}& \acs{WN}& \acs{PO}\\
		\hhline{~---||---}
		majority&\multicolumn{3}{c}{C-B-D} &\multicolumn{3}{c}{D-C-B}\\
		correct&2995 &2994 &2993 &2999 &2996 &3000\\
		C$\,\leftrightarrow\,$B&3 &4 &2 &1 &3 &0\\
		D$\,\leftrightarrow\,$B&2 &2 &5 &0 &1 &0\\
		\bottomrule
    	\end{tabular}
        \label{tab:permutation_comparison}
	\end{center}
    \vspace{-0.9cm}
\end{table}

For all datasets, our performance evaluation is based on two metrics that we calculate by averaging over all 3'000 test samples per set\footnote{Audio examples can be found at \url{https://www.audiolabs-erlangen.de/resources/2023-ICASSP-SSBRIPS}.}.
We report the \ac{SI-SDR} \cite{SISDR} of the reconstructed estimates and the \acf{SI-SDRi} over the mixture, namely both the mean over all regions and per region (denoted by, e.g., `driver' as indicated in Figure~\ref{fig:scatter-plot}).
The model comprises approximately $4.2$ million parameters, and training the models on an NVIDIA A100 SXM4 reserved only for this training took $109.5\,\mathrm{h}$ and $107.4\,\mathrm{h}$ for the variants with and without \ac{PIT}, respectively.
Further, as a multi-channel separation baseline where spatial information is not explicitly modeled, we trained an implementation of the \ac{MC ConvTasNet} \cite{MC-ConvTasNet} to maximize the \ac{SI-SDR} of the reconstructed signals in combination with \ac{PIT}.
Note that \ac{MC ConvTasNet} is not a task-specific baseline, i.e., it is not designed for region-based separation. The model comprises approximately $4.9$ million parameters, and training on an NVIDIA GeForce RTX2080 Ti took $53.1\,\mathrm{h}$.

Table \ref{tab:performance_comparison} shows the evaluation results for the three test sets.
For the \ac{FC} set, which matches the training condition, \ac{MC ConvTasNet} obtained an average \ac{SI-SDRi} of $16.8\,\text{dB}$, whereas \ac{SpaRSep} \cite{AmbiSep} obtained an average \ac{SI-SDRi} of $18.3\,\text{dB}$ for both training conditions, with \ac{PIT} and with the proposed fixed permutation at the output.
Note that \ac{SpaRSep} consistently outperforms \ac{MC ConvTasNet}, indicating that the triple-path structure is indeed advantageous for modeling spatial information.
The \ac{SI-SDRi} is approximately equal for all regions, which is also illustrated in Figure~\ref{fig:scatter-plot} for the \ac{SpaRSep} model trained with the proposed fixed mapping.
Each dot represents the \ac{SI-SDRi} per test point as described in Section~\ref{sec:ExSet}.
The mean \ac{SI-SDRi} values per point range between $15.9$ and $20.4$ decibels, represented by the diameter.
Since the input \ac{SI-SDR} is higher for regions D and C compared to region B, \iffalse accordingly,\fi the output \ac{SI-SDR} is also higher for regions D and C.
This difference stems from the mixing procedure, where the anechoic input signals are normalized and scaled as per the LibriMix definitions \cite{LibriMix}, but the reverberation introduces different direct-to-reverberation ratios for the spatial regions. 
% where the power ratios as defined for LibriMix \cite{LibriMix} were used for the sources before they were convolved with the \acp{RIR}, rather than determining the power ratios at the microphones after the convolution.
Further, even for the short reverberation times of the car scenario, the backseat signals have a lower direct-to-reverberation ratio and it is challenging to recover the reverberation tail.
% must be expected to have longer tails of late reverberation, which are challenging to reconstruct and thereby negatively influence the \ac{SI-SDR} values.

The performance is degraded by roughly $1.0\,\text{dB}$ for the \ac{WN} set, which is a condition not seen during training.
The model trained on noise-free data can thus be expected to generalize to real measurements, including microphone self-noise out of the box.
For the \ac{PO} set, we see that the model trained with fixed output mapping is superior to the \ac{PIT} model.
For both trained models, the metrics are higher by $4\,\text{-}\,5\,\text{dB}$ compared to the \ac{FC} set.
Since partial overlap is a more likely scenario in a natural human conversation than fully concurrent speakers, this is a welcome finding.

The performance similarity between the models with and without PIT suggested that the PIT model may have learned a fixed output mapping.
Table~\ref{tab:permutation_comparison} shows the different permutations obtained.
For $99.8\,\%$ of the examples (for the \ac{FC} set), the \ac{PIT} model output a common permutation of C-B-D. 
Interpreting this as a fixed permutation, we see that the confusions happen only between regions D$\,\leftrightarrow\,$B and C$\,\leftrightarrow\,$B.
The output permutation is found to be only slightly sensitive to the evaluation condition.
For all test sets, $\geq\!99.8\,\%$ of the samples were output in the `majority permutation'.

The fixed-mapping model generates output with the desired fixed permutation `D-C-B' most of the time ($\geq\!99.9\,\%$).
For the \ac{PO} set, there were no confusions at all, indicating that spatial information was properly learned.
Note that due to the random order of speaker onsets, this indicates that the assignment of speakers to regions also reliably works when only one source is active.
When confusion occurs for the fixed-mapping model, it is found to be between one of the front regions (D, C) and the back region (B).
Intuitively, this is expected since the possible ranges of \acp{DOA} of sources from the two front regions are non-overlapping, and the proposed architecture with fixed output mapping can generate region-specific outputs.
For each of the two front regions, respectively, and the back region, there exists a range of \acp{DOA} that is common for both regions.
In Figure~\ref{fig:scatter-plot}, these overlap regions are indicated by blue dotted lines merging at the array center.
Note that the distinguishability of two sources located in such an overlap region is influenced by their different direct-to-reverberation ratios.
Since random combinations of one point from every region were simulated, the mean \ac{SI-SDRi} does not degrade noticeably.

\begin{figure}[t]
    \centering
    \vspace{0.5em}
    \begin{tikzpicture}
    \node(fig1){\includegraphics[width=0.8\linewidth]{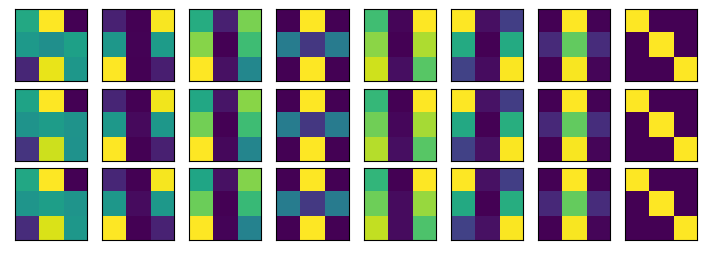}};
    \node[left = 0.35 cm of fig1](c1) {\rotatebox{90}{a) Trained with }};
    \node[below=0.0cm of fig1](fig2){
    \includegraphics[width=0.8\linewidth]{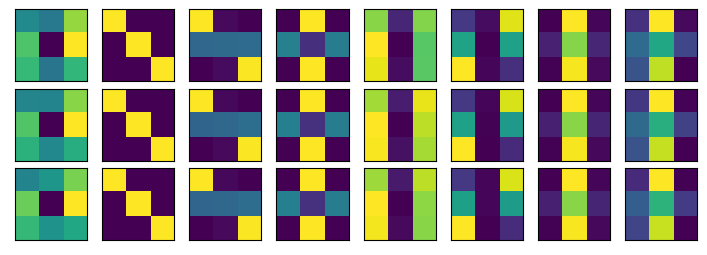}};
    \node[left = 0.35 cm of fig2](c2) {\rotatebox{90}{b) Trained with}};
        \node[left = 0.0 cm of fig2](c2) {\rotatebox{90}{PIT}};
              \node[left = 0.0 cm of fig1](c2) {\rotatebox{90}{ Fixed Mapping}};
    \end{tikzpicture}
    \vspace{-0.5em}
    \caption{Average attention weights computed by the $8$ heads (per row) of the inter-channel transformer in the first \ac{TPB}. The first row shows the attention weights when the source is active in region D and there is no source in the other regions. Similarly, the second and third rows correspond to sources in regions C and B. }
    \label{fig:attention_weights} 
    %\vspace{-0.5em}
\end{figure}
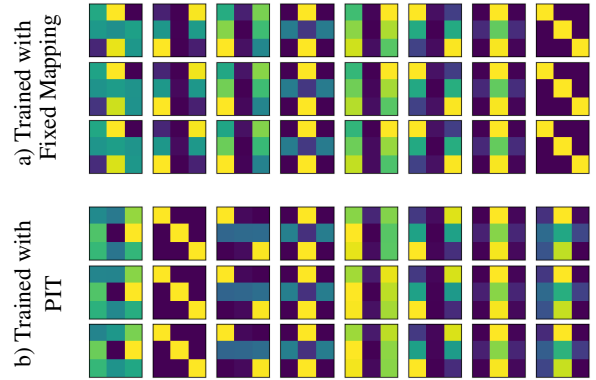

To interpret the representations computed by the separation network, and especially to probe the spatial processing discussed before (i.e., the inter-channel modeling), we studied the inter-channel attention for scenarios where only one source is active in one of the regions and the sources in the other regions are silent.
Figure~\ref{fig:attention_weights} shows the average attention weights ($M \times M$ matrices) computed in the eight heads of the inter-channel transformer of the first \ac{TPB}, which is also the first transformer layer in the network, directly after the encoder.
The attention weights are averaged over all time frames of $10$ different test samples.
We see that irrespective of the active source region, the average attention weights have a fixed pattern, i.e., the network computes representation based on fixed spatial processing in the first inter-channel transformer layer.
This behavior is observed for the model trained with fixed output mapping as well as the model trained with PIT.
Specific patterns computed by a few heads are also similar between the two models, e.g., head 8 of the fixed-order model and head 2 of the \ac{PIT} model.
We interpret this behavior as a sort of signal-independent beamforming, where the network learns a fixed spatial pre-processing of the encoded signals before signal-dependent separation is performed.
The inter-channel transformers in the other \acp{TPB} are found to compute active source region-specific attentions.

\section{\textbf{Conclusion}}
\label{sec:Conclusion}
We formulated the task of source separation that preserves region information. We proposed to address that task by a special training data generation scheme and by training with fixed correspondences of regions to separated sources on the network output. The employed transformer-based network architecture was shown to learn spatial features. We conclude that networks can be trained to separate sources and preserve regional information.

\iffalse
\section*{\centering{\normalsize{\MakeUppercase{Acknowledgment}}}}

The authors gratefully acknowledge the scientific support and HPC resources provided by the Erlangen National High Performance Computing Center (NHR@FAU) of the Friedrich-Alexander-Universit{\"a}t Erlangen-Nürnberg (FAU).
\newpage
\fi

\begin{acronym}[SI-SDR]
 \acro{DNN}{deep neural network}
 \acro{DOA}{direction-of-arrival}
 \acrodefplural{DOA}{directions-of-arrival}
 
 \acro{FC}{full concurrent test set}
 
 \acro{MC ConvTasNet}{multi-channel extension of ConvTasNet}
 \acro{MIMO}{multiple-input multiple-output}
 
 \acro{PIT}{permutation invariant training}
 \acro{PO}{test set with partial overlap}
 \acro{PReLU}{Parameterized \ac{ReLU}}
 
 \acro{ReLU}{rectified linear unit}
 \acro{RIR}{room impulse response}
 
 \acro{SI-SDR}{scale-invariant signal-to-distortion ratio}
 \acro{SI-SDRi}{improvement in \ac{SI-SDR}}
 \acro{SpaRSep}{separator network based on spatial regions}
 
 \acro{TF}{time-feature}
 \acro{TPB}{triple-path block}
 
 \acro{ULA}{uniform linear array}
 
 \acro{WN}{full concurrent with white noise test set}
\end{acronym}


\cleardoublepage

\balance
\bibliographystyle{IEEEbib}
\bibliography{bibliography}

\end{document}